\documentclass{aa}

\usepackage{subfigure}
 
\usepackage{txfonts}
\usepackage{natbib}
\bibpunct{(}{)}{;}{a}{}{,} 
\begin{document}  

\title{Exploring the internal rotation of the extremely low-mass He-core white dwarf GD$\,$278 with TESS asteroseismology}

\author{Leila M. Calcaferro\inst{1,2},
        Alejandro H. C\'orsico\inst{1,2},  
        Leandro G. Althaus\inst{1,2}, Isaac D. Lopez\inst{3} \and J.J. Hermes\inst{4}}  
\offprints{lcalcaferro@fcaglp.unlp.edu.ar} 

\institute{$^1$ Grupo  de Evoluci\'on  Estelar y  Pulsaciones,  Facultad de 
           Ciencias Astron\'omicas  y Geof\'{\i}sicas, Universidad
           Nacional de La Plata, Paseo del Bosque s/n, (1900) La
           Plata, Argentina\\   $^{2}$ Instituto de Astrof\'{\i}sica
           La Plata, CONICET-UNLP, Paseo  del Bosque s/n, (1900) La
           Plata,
           Argentina\\   
           $^{3}$ Department of Physics and Astronomy, Iowa State University, Ames, IA 50011, USA \\
           $^{4}$ Department of Astronomy, Boston University, 725 Commonwealth Ave., Boston, MA 02215, USA\\
           \email{lcalcaferro@fcaglp.unlp.edu.ar}}
\date{Received}  

\abstract{The advent of high-quality space-based photometry, brought about by missions like Kepler/K2 and TESS, makes it possible to unveil the fundamental parameters and properties of the interiors of white dwarf stars, and particularly extremely low-mass white dwarfs, by the tools of asteroseismology.}
{We present an exploration of the internal rotation of GD$\,$278, the first pulsating extremely low-mass white dwarf that shows rotational splittings within its periodogram.}
{We assess the theoretical frequency splittings expected for different rotation profiles, and compare them to the observed frequency splittings of GD$\,$278. To this aim, we employ an asteroseismological model representative of the pulsations of this star, obtained by using the {\tt LPCODE} stellar evolution code and the {\tt LP-PUL} nonradial pulsation code. We also derive a rotation profile that results from detailed evolutionary calculations carried out with the {\tt MESA} stellar evolution code, and use it to infer the expected theoretical frequency splittings.}
{We found that the best-fitting solution when assuming linear profiles for the rotation of GD$\,$278 leads to values of the angular velocity at the surface and the center that are only slightly differential, and still compatible with rigid rotation. Additionally, the values of the angular velocity at the surface and the center for the simple linear rotation profiles and for the rotation profile derived from evolutionary calculations are in very good agreement. Also, the resulting theoretical frequency splittings are compatible with the observed frequency splittings, in general, for both cases.}
{The results obtained from the different approaches followed in this work to derive the internal rotation of GD$\,$278 agree. The fact that they were obtained employing two independent stellar evolution codes gives our results robustness. Our results suggest only a marginally differential behavior for the internal rotation in GD$\,$278, and, considering the uncertainties involved, very compatible with the rigid case, as has been observed previously for white dwarfs and pre-white dwarfs. The rotation periods derived for this star are also in line with the values determined asteroseismologically for white dwarfs and pre-white dwarfs in general.}    

  \titlerunning{Exploring the internal rotation of the ELM WD GD$\,$278}   
\authorrunning{Calcaferro et al.}  
  \maketitle


\section{Introduction}  
\label{intro}  

Low-mass helium (He)-core white dwarf (WD) stars are characterized by stellar masses $M_{\star}\lesssim 0.45\ M_{\odot}$, and are likely the result of intense mass loss episodes in close-binary systems, before the occurrence of the core He flash during the red giant phase of low-mass stars, which would then be avoided \citep{2013A&A...557A..19A,2016A&A...595A..35I}. In this way, they would harbor He cores, at variance with average-mass ($M_{\star}\sim 0.6\ M_{\odot}$) WDs, supposed to have C/O cores. In particular, binary evolution is the most likely scenario for the so-called extremely low-mass (ELM) WDs, with $M_{\star}\lesssim 0.18-0.20 \ M_{\odot}$ \citep{2013A&A...557A..19A,2016A&A...595A..35I,2018ApJ...858...14S,2019ApJ...871..148L}. Detailed evolutionary calculations  \citep{2001MNRAS.323..471A,2013A&A...557A..19A,2016A&A...595A..35I} indicate that element diffusion significantly affects the evolution of low-mass He-core WDs, including ELMs, since this phenomenon would induce multiple H-shell flashes in progenitors of WDs with $M_{\star}\gtrsim 0.18-0.20\ M_{\odot}$ that consume most of their H content, leading to a quick evolution in their cooling tracks. In contrast, progenitors of WDs with $M_{\star}\lesssim 0.18-0.20 \ M_{\odot}$ are not expected to experience any H flashes and, in this way, can sustain stable H burning, which slows their WD evolution down.

Recently, numerous low-mass and ELM WDs have been detected \citep[see, e.g.][]{2009A&A...505..441K,2010ApJ...723.1072B,2016ApJ...818..155B,2020ApJ...889...49B,2022ApJ...933...94B,2011ApJ...727....3K,2015ApJ...812..167G,2020ApJ...894...53K}. Interestingly, multi-periodic brightness variations have been measured in some low-mass and ELM WDs \citep[see e.g.][]{2012ApJ...750L..28H,2013ApJ...765..102H,2013MNRAS.436.3573H,2018MNRAS.479.1267K,2018A&A...617A...6B,2018MNRAS.478..867P,2021ApJ...922..220L}, giving rise to the new type of variable star, known as ELMVs. In addition, pulsations have also been detected in stars considered to be the low-mass WDs probable precursors \citep{2013Natur.498..463M,2014MNRAS.444..208M,2016ApJ...822L..27G,2020ApJ...888...49W}, known as low-mass pre-WDs, which opens up the opportunity of studying the evolution of the WD progenitors, and also defines another type of variable star, the pre-ELMVs. The existence of ELMVs and pre-ELMVs provides researchers a unique opportunity for probing the interiors of these variable stars by employing the tools of asteroseismology \citep{2008ARA&A..46..157W,2008PASP..120.1043F,2010A&ARv..18..471A,2019A&ARv..27....7C}. This has successfully been done for many pulsating WDs and pre-WDs such as ZZ Ceti stars or DAVs \citep[with H-rich atmospheres; see e.g.][]{2001ApJ...552..326B,2012MNRAS.420.1462R,2017A&A...598A.109G}, V777 Her stars or DBVs \citep[with He-rich atmospheres; see e.g.][]{2012A&A...541A..42C,2022A&A...659A..30C,2014A&A...570A.116B,2019A&A...632A..42B}, and pulsating PG 1159 or GW Vir stars \citep[hot pre-WD, with C-, O- and He-rich atmospheres;][]{2007A&A...461.1095C,2007A&A...475..619C,2008A&A...478..869C,2009A&A...499..257C,2021A&A...645A.117C,2016A&A...589A..40C,2022MNRAS.513.2285U}. In addition, \cite{2017A&A...607A..33C,2018A&A...620A.196C} have performed detailed asteroseismological analyses of all known (and alleged) pulsating ELMVs on the basis of a complete set of fully evolutionary models that represents low-mass He-core WD stars \citep{2013A&A...557A..19A,2018A&A...614A..49C}.
   
The pulsations observed in ELMVs are compatible with gravity ($g$) modes, and stability computations \citep{2012A&A...541A..42C,2013ApJ...762...57V,2016A&A...585A...1C} suggest that these modes are probably excited by the $\kappa-\gamma$ mechanism \citep{1989nos..book.....U} acting at the H-ionization region. In pre-ELMVs, the observed pulsations are compatible with radial and nonradial pressure ($p$) and (possibly) $g$ modes, and in this case, nonadiabatic analyses \citep{2013MNRAS.435..885J,2016A&A...588A..74C,2016ApJ...822L..27G}  indicate that the excitation mechanism may be the $\kappa-\gamma$ mechanism acting mainly at the zone of the second partial ionization of He, with a weaker contribution from the region of the first partial ionization of He and the partial ionization of H. In this way, the presence of He in the driving zone is necessary for the modes to be destabilized by the $\kappa-\gamma$ mechanism \citep{2016A&A...588A..74C}. As the presence of He in the driving zone would not be possible when the effects of element diffusion are taken into account (particularly, the gravitational settling, which would otherwise quickly deplete the region of He), a mechanism capable of preventing (or diminishing) its effects is needed \citep{2016A&A...588A..74C}. As \cite{2016A&A...595L..12I} showed, rotational mixing can, in principle, counteract the gravitational settling effects, maintaining enough He within the driving region of pre-ELMs, and then, allowing He to drive pulsational instabilities with frequencies in agreement with the observed ones. Further, the presence of metals such as calcium in the spectra of some ELMs \citep{2014ApJ...781..104G, 2014MNRAS.444.1674H}, also requires a gravitational settling-counteracting mechanism, which could also be rotational mixing, as \cite{2016A&A...595L..12I} calculations evidenced. Based on this scenario, if ELMs descend from pre-ELMs, then it is expected that pulsating ELMs are rotating. This being the case, it should be possible to detect the stellar rotation by the splitting of the pulsation frequencies, something that, until now, had been elusive \citep{2019A&ARv..27....7C}.

Rotation is a fundamental property of stars that have effects on their structure and evolution. Recently, the study of the internal rotation in pulsating stars from the main sequence to late stages in stellar evolution has been significantly boosted by asteroseismology \cite[see e.g.][]{2012ApJ...756...19D,2014ApJ...788...93C,2014MNRAS.444..102K,2021RvMP...93a5001A}. In nonrotating spherical stars, nonradial $g$ modes, which are characterized by a harmonic degree $\ell$ and a radial order $k$, are ($2\ell+1$)-fold degenerated with respect to the azimuthal order $m$. When rotation is present, that mode degeneracy is removed and hence, each pulsation frequency is split into multiplets of $2\ \ell + 1$ frequencies. The separations of the components of the multiplets are known as rotational splittings and are related to the angular velocity, allowing the estimation of the rotation period. The presence of multiplets helps also to identify the harmonic degree of the pulsation modes of the pulsating star.

The study of the internal rotation by means of asteroseismological tools in pulsating WDs and pre-WDs, for instance, by \cite{1999ApJ...516..349K,2009Natur.461..501C,2011MNRAS.418.2519C,2016ApJS..223...10G,2017ApJ...835..277H}, is of fundamental interest for the present work. In the case of GW Vir stars, which can be probed deeply by virtue of the behavior of the pulsation modes, \cite{1999ApJ...516..349K} made use of asteroseismological inversion techniques, analogous to the ones applied in helioseismology, to study their internal rotation profiles. They found that PG 1159-035 (prototype of GW Vir stars), may be rotating with differential rotation, although their results are inconclusive. \cite{2009Natur.461..501C} also studied the rotation in PG 1159-035, but with a forward-method approach, and concluded that this star may be rotating as a solid body. \cite{2011MNRAS.418.2519C} applied both forward and inverse methods on another GW Vir star, PG 0122+200, and concluded that its observed frequency splittings are consistent with a differential rotation profile, although rigid body rotation cannot be discarded. For ZZ Ceti stars, which can be probed less deeply, the work of \cite{2016ApJS..223...10G} for Ross 548 and GD 165, indicates that these stars may be rotating as solid bodies. The detailed and complete analysis that \cite{2017ApJ...835..277H} performed over a large set of isolated ZZ Ceti stars that show rotational splittings, allowed them to find a mean rotation period of $35\ $h for WDs with masses in the range of $0.51-0.73\ M_{\odot}$ \citep[although ZZ Ceti stars with faster rotation rates have also been found; see Table 10 of][]{2019A&ARv..27....7C}.  In general, the rotation periods for WD and pre-WD stars are between $\sim 1\ $h and $\sim 18\ $d \citep[][]
{2015ASPC..493...65K,2019A&ARv..27....7C}

In this work, we apply for the first time asteroseismological methods to probe the internal rotation of a pulsating ELM WD star, GD$\,$278, on the basis of the first measurement of rotational splittings in this type of WD stars \citep{2021ApJ...922..220L}. GD$\,$278 is a pulsating ELM WD from the \citet{2019MNRAS.482.4570G} Gaia DR2 catalog ($G$ = $14.9\ $mag), in a $4.61\ $h single-lined spectroscopic binary  \citep{2020ApJ...889...49B}, with a parallax distance determined from the Gaia eDR3 of $151.19 \pm 0.78\ $pc \citep{2020yCat.1350....0G,2021AJ....161..147B}. It is characterized by $T_{\rm eff}= 9230 \pm 100\ $K and $\log(g)= 6.627 \pm 0.056\ $[CGS], and its pulsation periods range from $\sim 2290$ to $6730\ $s \citep{2021ApJ...922..220L}.
This paper is organized as follows. In Sect.~\ref{evolutionary}, we give a brief summary of the numerical codes employed, while in Sect.~\ref{data_properties}, we provide the observational data and the properties of the stellar models utilized. In Sect.~\ref{analysis}, we carry out rotational splitting fits comparing the observed and the theoretical frequency splittings (Sect.~\ref{fits}), the latter obtained by considering different types of rotation profiles. Next, we employ a rotation profile resulting from stellar evolutionary calculations that characterize the stars under study, and use it to derive the predicted values of the theoretical frequency splittings (Sect.~\ref{from_evol}). A comparison of the results is displayed in Sect.~\ref{compare-results}. Finally, in Sect.~\ref{conclusions},  we summarize our main findings.

\section{Evolutionary models}  
\label{evolutionary}  

In this paper we employed fully evolutionary models of
low-mass He-core WDs generated with two independent stellar evolution codes, the {\tt LPCODE} \citep{2005A&A...435..631A,
2009A&A...502..207A,2013A&A...557A..19A,2015A&A...576A...9A} and the {\tt MESA} 
\citep[Modules for Experiments in Stellar Astrophysics,][]{Paxton2011, Paxton2013, Paxton2015, Paxton2018, Paxton2019} code. 
Both compute in detail the complete
evolutionary stages that lead to the WD formation, allowing for the study
of the WD evolution consistently  with the predictions of the
evolutionary  history of progenitors. 
We briefly mention  the main  ingredients employed for  our analysis  of  low-mass He-core WDs.

Realistic configurations of low-mass He-core WDs resulting from the {\tt LPCODE} stellar evolution code were computed from detailed evolutionary calculations that mimic the binary evolution of the progenitor stars \citep[see][for details]{2013A&A...557A..19A}.  Binary evolution was assumed to be fully nonconservative, and the losses of angular momentum due to mass loss, gravitational wave radiation, and magnetic braking were considered. Binary configurations assumed in \cite{2013A&A...557A..19A} consist of an evolving main-sequence low-mass component (donor star) of initially $1 M_{\sun}$, and a $1.4 M_{\sun}$ neutron star companion as the other component. The metallicity of the progenitor stars is $Z = 0.01$. Initial He-core WD models with stellar masses ranging from $0.1554$ to $0.4352\ M_{\sun}$ were derived from stable mass loss via Roche-lobe overflow. The evolution of these models was computed down to the range of luminosities of cool WDs, including the stages of multiple thermonuclear CNO flashes at the beginning of the cooling branch when appropriate. We also considered time-dependent diffusion due to gravitational  settling and chemical  and thermal diffusion of nuclear  species  following  the multicomponent  gas  treatment  of \citet{1969fecg.book.....B}. Further details can be found in \cite{2013A&A...557A..19A}. Adiabatic pulsation periods for nonradial dipole and quadrupole ($\ell=1, 2$, respectively) $g$ modes were computed employing the adiabatic version of the LP-PUL pulsation code \citep{2006A&A...454..863C,2009ApJ...701.1008C}.

Models considering rotational mixing and element diffusion were generated employing version 11701 of the {\tt MESA} stellar evolution code \citep{Paxton2011, Paxton2013, Paxton2015, Paxton2018, Paxton2019}, following the binary evolution of the progenitor stars from the ZAMS. The donor has an initial stellar mass of $1 M_{\sun}$, and its point-mass (neutron star) companion, $1.4 M_{\sun}$. The evolution was computed in detail throughout the mass transfer episodes via stable Roche-lobe overflow until the WD cooling branch. The mass transfer proceeded via the Ritter scheme \citep{Ritter1988}, and took place while the donor overfills its Roche lobe. We considered that $30\%$ of the transferred mass is lost from the vicinity of the neutron star as fast wind. Initial He-core WD models with $M_{\star}$ in the range of $0.17 - 0.3\ M_{\sun}$ were derived in this way. The chemical and angular momentum transport resulting from rotationally induced instabilities is implemented in a diffusive manner in MESA \citep{2000ApJ...528..368H,Paxton2013,Paxton2019}. We included the effects of rotationally induced mixing processes, following the \citet{2016A&A...595A..35I} scheme, taking into account the mixing due to secular shear instability and dynamical shear instability, Goldreich-Schubert-Fricke instability, and Eddington-Sweet circulation \citep{2000ApJ...528..368H,2005ApJ...626..350H}. Also considered are the mixing of angular momentum in radiative regions because of dynamo-generated magnetic fields \citep{2002A&A...381..923S} and the transport of angular momentum given by electron viscosity \citep{1987ApJ...317..733I}. Following \citet{2016A&A...595A..35I}, we adopt for the ratio of the turbulent viscosity to the diffusion coefficient the value $f_c= 1/30$, and $f_{\mu} = 0.1$ for the sensitivity to compositional gradients \citep[see][for further details]{2000ApJ...528..368H,Paxton2013,2016A&A...595A..35I}. In our calculations, the initial rotation velocity is set such that the star is synchronized with the initial orbital period, and the effect of tides and spin-orbit coupling are taken into account.

\section{The data and the asteroseismological model}
\label{data_properties}

GD$\,$278 is a pulsating ELM WD in a $4.61\ $h single-lined binary  \citep{2019MNRAS.482.4570G,2020ApJ...889...49B,2021ApJ...922..220L}. The pulsations were first detected via photometry obtained with the Otto Struve telescope at McDonald Observatory and later characterized using Transiting Exoplanet Survey Satellite \citep[TESS,][]{2015JATIS...1a4003R} observations, which revealed the richest set of periods of a pulsating ELM \citep{2021ApJ...922..220L}. Its spectroscopic parameters are $T_{\rm eff}= 9230 \pm 100\ $K and $\log(g)= 6.627 \pm 0.056\ $[CGS], corresponding to a stellar mass of
$M_{\star}= 0.191  \pm 0.013\ M_{\odot}$ \citep{2020ApJ...889...49B}.

A thorough mode identification of the pulsation frequencies allowed \citet{2021ApJ...922..220L} to  distinguish eight possible rotational splittings associated with $\ell= 1$ and $\ell= 2$, as can be seen in Table~\ref{tab:observ} \citep[adapted from][ see their Table 2]{2021ApJ...922..220L}, the first time that such an effect is detected in an ELM WD. In the table, the first two columns indicate the harmonic degree and the azimuthal index of the modes. The third and fourth columns show the observed pulsation periods and corresponding frequencies, respectively. The fifth column indicates the uncertainty of the observed frequencies. The sixth one gives the observed frequency splittings, and the last one indicates their uncertainties.

Employing the mode identification from Table~\ref{tab:observ}, we compared the observed periods associated with $m= 0$ (that is, the central components of the rotational splittings), to the theoretical periods of the set of low-mass He-core WD models generated with the {\tt LPCODE} stellar evolution code, with the aim of searching for a model that represents a good global match to the periods observed in GD$\,$278, and that also lies within the constraints given by the spectroscopic parameters for this star. This period-to-period fit is analogous to the procedure followed by \citet{2017A&A...607A..33C,2018A&A...620A.196C} for other ELMVs. In this way, we adopt an asteroseismological model\footnote{We note that this is a different asteroseismological model than that adopted in \citet{2021ApJ...922..220L}. In that work, we allowed for different H envelopes in our asteroseismological fits. However, given the exploratory nature of this work, we decided to use a model harboring a canonical (thick) envelope, for the sake of simplicity.} characterized by $M_{\star}= 0.186\ M_{\odot}$, $T_{\rm eff}= 9335\ $K and $\log(g)= 6.618\ $[CGS]. For this model, we show in Table~\ref{tab:o} the radial order and harmonic degree (first and second columns), the observed and theoretical periods (third and fourth columns), and the relative difference between the last two quantities (fifth column).

\begin{table*}
\centering
\caption{Mode identifications and observed frequency splittings for GD$\,$278 \citep[adapted from][]{2021ApJ...922..220L}.} 
\begin{tabular}{crcccccc}
\hline
  $\ell$  & $m$  & $\Pi^O$ [s]  & $\nu^O$ [$\mu$Hz] & $\sigma_{\nu^O}$  [$\mu$Hz]& $\delta \nu^O$ [$\mu$Hz] & $\sigma_{\delta \nu^O}$  [$\mu$Hz]\\
\hline
1	&	$-$1	&	2447.86	&	408.52	&	0.03	&	\\
1   &     0     &   2367.4   &            &               \\
1	&  	 1	   &	2292.0	&	436.29	&	0.06	&	2$\times$13.89	& 2$\times$0.09	 \vspace{1mm}\\

1	&	0	&	2658.51 	&	376.15	&	0.04		& \\
1	&	+1	&	2569.0 	&	389.26		&		0.06	&		13.10		&		0.10 \vspace{1mm}	\\

1	&	$-$1	&	3365.56		&	297.127		&	0.012	&					\\
1	&	0	&	3226.1	 	&	309.98	& 	0.06		&	12.85	&	0.07 \vspace{1mm}	\\

1	&	$-$1	&	4259.1		&	234.79	&	0.03	&					\\
1	&	0	&	4028.83		&	248.211	&	0.013	&		13.42 &		0.04	\\
1	&	1	&	3815.8		&	262.07	&	0.05	&		13.85	&	0.06	\vspace{1mm}\\

1	&	$-$1	&	5135.1	 	&	194.74	&	0.03	& \\
1	&	0	&	4756.5	 	&	210.24	&	0.03		&	15.50	&	0.06	\\
1	&	1	&	4428.5	 	&	225.812	&	0.028		&	15.57	&	0.06	\vspace{1.4mm}\\

2	&	$-$2	&	6295.4 	&	158.846	&	0.026	&\\
2	&	$-$1	&	5550.6	 	&	180.161	&	0.025	&		21.32	&	0.05	\\
2	&	0	&	4959.6	 	&	201.63	&	0.04		&	21.46	&	0.07	\\
2	&	1	&	4473.1		&	223.56	&	0.05		&	21.93	&	0.09	\\
2	&	2	&	4083.5	 	&	244.89	&	0.04	&		21.33 &		0.09	\vspace{1mm}\\

2	&	$-$1	&	5814.8		&	171.975	&	0.021	&				\\
2	&	0	&	5198.3	&	192.370	&	0.029		&	20.39&		0.05	\\
2	&	+1	&	4628.8	 	&	216.039	&	0.022	&		23.67&		0.05	\vspace{1mm}\\

2	&	$-$2	&	6729.0		&	148.61	&	0.04	&				\\
2	&	$-$1	&	5896.8	 	&	169.58	&	0.03	&		20.97	&	0.07	\\
2	&	0	&	5253.2	&	190.359	&	0.007	&		20.78 &		0.04	\\
2	&	2	&	4270.8		&	234.15	&	0.05	&		2$\times$21.89	&	2$\times$ 0.05	\\
\hline
\end{tabular}
\label{tab:observ}
\end{table*}

\begin{table}
\centering
\caption{Comparison between the observed periods of GD$\,$278 and the theoretical periods of the adopted asteroseismological model characterized by $M_{\star}= 0.186\ M_{\odot}$, $T_{\rm eff}= 9335\ $K and $\log(g)= 6.618\ $[CGS]. This stellar model has undergone CNO flashes during its early-cooling phase \citep[][]{2013A&A...557A..19A}.}
\begin{tabular}{ccccc}
\hline
$k$ &  $\ell$  & $\Pi^O$ [s] & $\Pi^T$ [s] & $|\Delta \Pi|/\Pi$\\
\hline
25  &   1   &2367.4    & 2332.6513  & 0.0147               \\
29	&	1	&2658.51   & 2681.3868  & 0.0086\\
35	&	1	&3226.1	 	&3212.6071	& 0.0042\\
44	&	1	&4028.83	&4017.2143	&0.0029	\\
52	&	1	&4756.5	 	&4735.2063	&0.0045\vspace{1mm}	\\
94	&	2	&4959.6	 	&4940.8258	&0.0038\\
99	&	2	&5198.3	&5203.3066	&0.0010\\
100	&	2	&5253.2	&5255.5214	&0.0004\\
\hline
\end{tabular}
\label{tab:o}
\end{table}

\subsection{Rotational splittings}
As mentioned before, rotation lifts the $g$ mode degeneracy of a nonrotating stellar structure, splitting each pulsation frequency into multiplets of $2\ell+1$ frequencies for different values of $m$ \citep{1989nos..book.....U}. Considering that the star is a slow rotator, each multiplet component is given by (to first order):

\begin{equation}
\nu_{k\ell m}(\Omega)= \nu_{k\ell}(\Omega=0) + \delta \nu_{k\ell m},
\end{equation}

\noindent where $\nu_{k\ell m}(\Omega)$ is the frequency of the mode with indexes ($k,\ell,m$), $\nu_{k\ell}(\Omega=0)$ is the frequency of the degenerate mode with indexes ($k,\ell$) without rotation, and $\delta \nu_{k\ell m}$ is the rotational splitting which, 
assuming rigid rotation (constant $\Omega$), is given by:

\begin{equation}
\label{delta-nu-ckl}
\delta \nu_{k\ell m}= -m \frac{\Omega}{2 \pi} (1-C_{k\ell}),
\end{equation}

\noindent with $m= 0, \pm 1, ..., \pm \ell$ and $\Omega$, the angular velocity. The $C_{k\ell}$ are coefficients that depend on the stellar structure details as \citep{1949ApJ...109..149C,1951ApJ...114..373L}:

\begin{equation}
\label{ckl}
C_{k\ell}= \frac{\int^{R_*}_0{\rho r^2 [2 \xi_r \xi_t + \xi^2_t] dr}}{\int^{R_*}_0{\rho r^2 [\xi^2_r + \ell(\ell + 1) \xi^2_t] dr}},
\end{equation}

\noindent where $\xi_r$ y $\xi_t$ are the unperturbed radial and tangential eigenfunctions, respectively. These  eigenfunctions are obtained from the nonrotating case. Note that for $g$ modes, and for large values of $k$, $\xi_r \ll \xi_t$, so that $C_{k\ell} \rightarrow 1/ \ell ( \ell+1)$ \citep{1975MNRAS.170..405B}.

When the rigid body condition is relaxed and differential rotation is allowed, considering spherical symmetry, that is, $\Omega= \Omega(r)$, the frequency splittings are calculated as: 

\begin{equation}
\label{delta_nu_t_ker}
\delta \nu_{k\ell m}= -m \int^{R_{\star}}_{0} \frac{\Omega (r)}{2 \pi} K_{k\ell}(r) dr,
\end{equation}

\noindent where $K_{k\ell}(r)$ are the first-order rotational kernels obtained from the rotationally unperturbed eigenfunctions as:

\begin{equation}
\label{kernel}
K_{k\ell}(r)= \frac{\rho r^2 \{\xi_r^2 -2\xi_r\xi_t - \xi^2_t[\ell(\ell + 1)-1]\}}{\int^{R_*}_0{\rho r^2 [\xi^2_r + \ell(\ell + 1) \xi^2_t] dr}},
\end{equation}

\noindent which are then determined by the stellar model. 

\begin{figure}  
\centering   \includegraphics[clip,width=250pt]{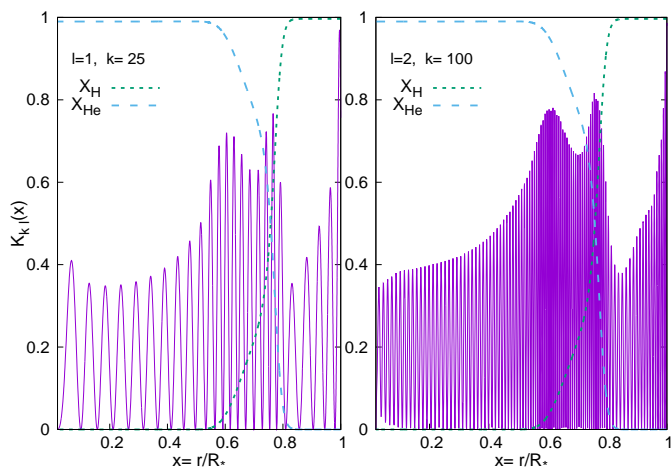} 
\caption{Normalized rotational kernels $K_{k\ell}(r)$ for $\ell= 1$, $k= 25$ (\textit{left panel}) and $\ell= 2$, $k= 100$ (\textit{right panel}). Also indicated are the chemical profiles of H (green dashed lines) and He (light-blue dashed lines).}
\label{fig:kernels}
\end{figure}

Eq.~(\ref{delta_nu_t_ker}) indicates that the sensitivity of the pulsation modes to the internal rotation can be analyzed from the behavior of the rotational kernels for each mode. In Fig.~\ref{fig:kernels} we show the normalized rotational kernels computed from our adopted asteroseismological model for GD$\,$278 for the shortest period with $\ell= 1$, $k= 25$ (left panel), and the longest period with $\ell= 2$, $k= 100$ (right panel). 
It is clear from the figure that the rotational kernels of these modes have the largest amplitudes in the region $0.5 \lesssim r/R_* \lesssim 0.8$ and in the outer part of the models, but they also have appreciable amplitudes throughout the entire stellar model. This suggests that the $g$ modes observed in GD$\,$278 are sensitive to the whole rotation profile, similarly to the behavior of the rotational kernels in GW Vir stars, and at variance with the case of ZZ Ceti and V777 Her stars that only probe the outer parts of the star \citep{1999ApJ...516..349K,2011MNRAS.418.2519C,2016ApJS..223...10G}.

\section{Analysis and results}  
\label{analysis}

\subsection{Rotational splittings fits}
\label{fits}
To start with this analysis, we compare the observed frequency splittings, $\delta \nu^{\rm O}$, to the theoretical frequency splittings, $\delta \nu^{\rm T}$, which are calculated by varying the rotation profile, and employing the asteroseismological model presented in the previous section. This comparison is done by estimating the goodness of the match between these quantities as \citep{2009Natur.461..501C,2011MNRAS.418.2519C}:

\begin{equation}
\label{quality}
\chi^2=  \frac{1}{n} \sum_{i=1}^n  \frac{1}{\sigma_{\delta \nu^{\rm O}}^2} (\delta \nu_i^{\rm   O} -   \delta \nu_i^{\rm  T})^2, 
\end{equation}

\noindent where the index $i$ identifies each splitting, and $n$, the number of rotational splittings considered. In this expression, each term is weighted with the inverse of the squared uncertainty of the observed frequency splittings ($\sigma_{\delta \nu^{\rm O}}^2$), calculated from Table~\ref{tab:observ} \citep{2021ApJ...922..220L}.

\subsubsection{Rigid rotation}
\label{rigid-profile}
\begin{figure}  
\centering   \includegraphics[clip,width=250pt]{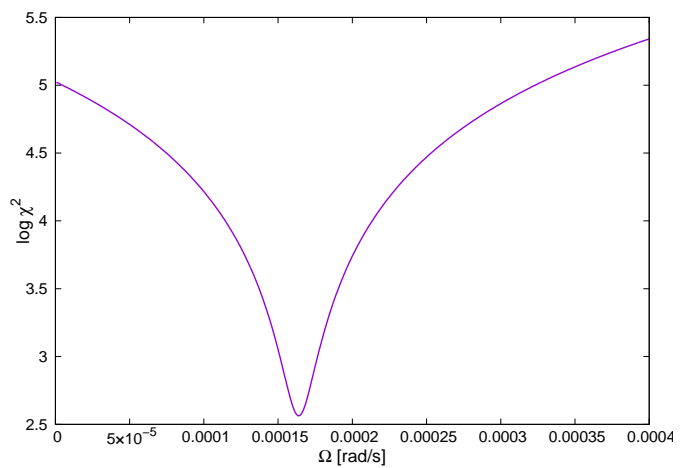} 
\caption{Logarithm of the quality function $\chi^2$ in terms of $\Omega$, assuming rigid body rotation. For this calculation, we have considered all the multiplets associated with $\ell= 1, 2$ from Table~\ref{tab:observ}.}
\label{fig:rigid}
\end{figure}

First, we consider the simplest case for the rotation profile, which is rigid body rotation, i.e. the angular velocity is constant throughout the star. We consider values of $\Omega/2 \pi$ from $1 \times 10^{-7}\ $ Hz to $1 \times 10^{-4}\ $Hz (corresponding to periods of $\sim 115\ $d and $3\ $h, respectively),  and for each one we assess the theoretical frequency splittings according to Eq.~(\ref{delta-nu-ckl}), where the coefficients $C_{k\ell}$ are determined for each mode by using Eq.~(\ref{ckl}). Considering all the observed frequency splittings associated with $\ell= 1,2$ from Table~\ref{tab:observ}, we search for the best match between the observed and the theoretical frequency splittings by calculating $\chi^2$ (Eq.~(\ref{quality})). We obtain the results shown in Fig.~\ref{fig:rigid}, where we plot the logarithm of the quality function $\chi^2$ in terms of $\Omega$. In the figure, the minimum (that is, the best fit) indicates a well-defined solution for $\Omega \sim 1.6 \times 10^{-4}\ $rad s$^{-1}$, corresponding to a rotation period $\sim 10.9\ $h. This value of the rotation period of GD$\,$278 is in line with the determination of the rotation periods for other WDs and pre-WDs \citep{2015ASPC..493...65K,2019A&ARv..27....7C}.

\subsubsection{Differential rotation: the linear case}
\label{linear}
Here we lift the solid-body rotation condition, and, given the exploratory nature of this work, allow for a simple case of linear differential rotation:

\begin{equation}
\label{eq:linear}
\Omega(r)= \Omega_C + (\Omega_S-\Omega_C)r
\end{equation}

\noindent where $\Omega_C$ and $\Omega_S$ indicate the rotation velocities at the center and at the surface, respectively. This set of linear profiles comprises both cases of increasing and decreasing $\Omega(r)$ with $r$, and also the rigid-body rotation case ($\Omega_C= \Omega_S$), already shown in the previous section.

\begin{figure}  
\centering   \includegraphics[clip,width=250pt]{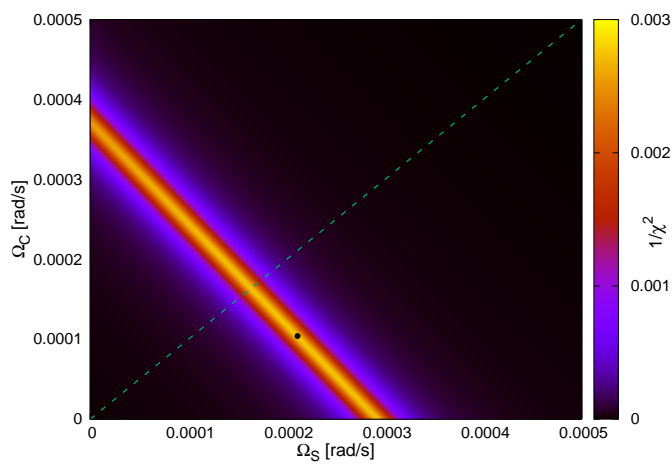} 
\caption{Contour map of the inverse of the quality function $1/\chi^2$ showing the goodness of the fits in the plane $\Omega_S - \Omega_C$, considering all the observed rotational splittings associated with $\ell= 1, 2$. We consider that GD$\,$278 rotates following a linear profile. The green dashed line indicates the corresponding solutions according to rigid rotation. The best-fitting solution is indicated with a black dot.}
\label{fig:18splitt}
\end{figure}

For the calculation of the quality function, via Eq.~(\ref{quality}), this time we compute the theoretical frequency splittings using Eq.~(\ref{delta_nu_t_ker}), with the rotational kernels determined by Eq.~(\ref{kernel}). We first consider all the observed frequency splittings associated with $\ell= 1, 2$ from Table~\ref{tab:observ}. We show the results in Fig.~\ref{fig:18splitt}, where we plot the inverse of the quality function, $1/\chi^2$, in the $\Omega_S - \Omega_C$ plane, along with a green dashed line indicating the loci of the solutions considering rigid rotation. By plotting the inverse of the quality function, the higher its value, the better the match between the observed and theoretical frequency splittings. It is clear from the figure that there is a range of possible solutions, that is, any solution lying on the yellow zone would have almost the same quality, therefore suggesting that the rotation could be rigid as well as differential. We indicate in the figure the best-fitting solution with a black dot, corresponding to ($\Omega_S,\Omega_C) \sim (2.1 \times 10^{-4}, 1.0 \times 10^{-4}$) rad s$^{-1}$. This solution, that lies close to the rigid rotation line, formally corresponds to differential rotation. This suggests the possibility that the stellar surface may be rotating slightly faster than the stellar center.

It is interesting to study the sensitivity of our results to each of the observed frequency splittings, for which we carry out a simple procedure: we alternately remove each individual multiplet and evaluate the corresponding solution. When doing so, we find that in general, the results remain comparable between each other, that is, in average ($\Omega_S,\Omega_C)\sim (2.0 \times 10^{-4}, 1.0 \times 10^{-4}$) rad s$^{-1}$. However, for three cases, the results change considerably. For instance, when we exclude from the dataset the multiplet associated with $k= 100$, we find that $\Omega_S$ becomes significantly smaller ($\sim 6 \times 10^{-7}\ $rad s$^{-1}$), implying a rotation period extremely long, which would suggest that it is crucial to include this multiplet in the calculations. Its inclusion may also be justified by noticing that the frequency of the central component of this multiplet has the largest amplitude in the TESS periodogram \citep[see Table 1 from][]{2021ApJ...922..220L}. A similar situation occurs when we exclude the multiplets associated with $k= 94$ and $52$ from the analysis.
 
It is worth mentioning that we also tried other types of rotation profiles, for instance, two-zone profiles as presented by \cite{2009Natur.461..501C} for PG 1159-035, and also  piecewise functions with three parts. The former does not allow us to make any meaningful conclusions of the rotation properties for GD$\,$278, indicating that this type of profiles may not be adequate for modeling the rotation in GD$\,$278, while the latter leads to almost the same results as with the linear simple profiles of Eq.~(\ref{eq:linear}).

\subsection{Rotation profile resulting from the evolution}
\label{from_evol}
Here we follow a different approach by employing a profile for the angular velocity $\Omega(r)$ that results from detailed evolution calculations\footnote{We are aware of the existing uncertainties involved in the rotational mixing processes, such as the effects of the Eddington-Sweet circulation, and in the transport of angular momentum, which translate into uncertainties in the rotation profile. However, given the exploratory nature of this work, we believe that a physical rotation profile calculated with MESA, such as the one used here, is worth considering. A detailed discussion of the uncertainties involved in the treatment of rotation in MESA is out of the scope of the present paper.}, instead of assuming an artificial, simple profile as in the previous sections. By using the calculated $\Omega(r)$ profile, we estimate the corresponding values of the theoretical frequency splittings.

\begin{figure}  
\centering   \includegraphics[clip,width=250pt]{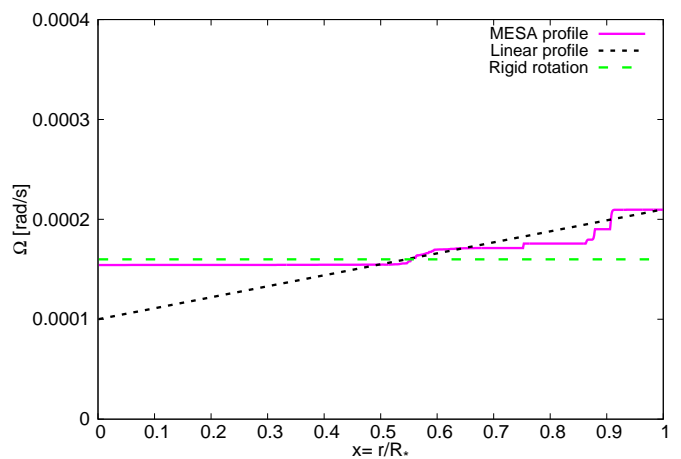} 
\caption{Rotation profile of the {\tt MESA} model with $M_{\star}= 0.186\ M_{\odot}$, $T_{\rm eff}= 9335\ K$ and log(g)= 6.618 [CGS] (magenta solid line), as resulting from our evolutionary calculations. Also included are the rigid (green dashed line) and linear (black dotted line) rotation profiles (see text for details)}.
\label{fig:omega-mesa}
\end{figure}

For this procedure we adopt a low-mass He-core model generated with {\tt MESA} (see Sect.~\ref{evolutionary}), characterized by $M_{\star}= 0.186\ M_{\odot}$, $T_{\rm eff}= 9335\ K$, and $\log(g)$= 6.618 [CGS], that \textit{closely} matches our asteroseismological model  (that is, they have similar values of $M_{\star}$, $T_{\rm eff}$, and $\log (g)$, along with other parameters such as the stellar radius, the H-envelope content, etc.). The rotation profile of this model is shown in Fig.~\ref{fig:omega-mesa} (magenta solid line), and is characterized by ($\Omega_S,\Omega_C$) $\sim$ ($2.1 \times 10^{-4}, 1.5 \times 10^{-4}$) rad s$^{-1}$. For illustrative purposes, we also include the rigid rotation profile obtained in Sect.~\ref{rigid-profile} (green dashed line), and the linear rotation profile of the best-fit model obtained in Sect.~\ref{linear} (black dotted line).
The {\tt MESA} rotation profile and its features, which show that the rotation is barely differential and close to uniform, with the surface rotating slightly faster than the center, are the result of the CNO flashes that the progenitor of this WD experienced before entering its final cooling track, as \cite{2016A&A...595A..35I} shows (see their Section 4.1.1.).

Employing the $\Omega(r)$ profile of this model to calculate the theoretical frequency splittings corresponding to the modes of interest, by means of Eq.~(\ref{delta_nu_t_ker}), we obtain the results shown in Table~\ref{tab:compar-splitt}. In the table, the first column indicates the radial order of the modes, the second and third columns represent the theoretical and observed frequency splittings, respectively, while the last column shows the relative difference between these quantities. The table indicates that, in general, there are discrepancies between the observed and the theoretical frequency splittings, but the relative differences remain below $\sim 15\%$.

\subsection{Comparison between results}
\label{compare-results}
\begin{figure}  
\centering   \includegraphics[clip,width=250pt]{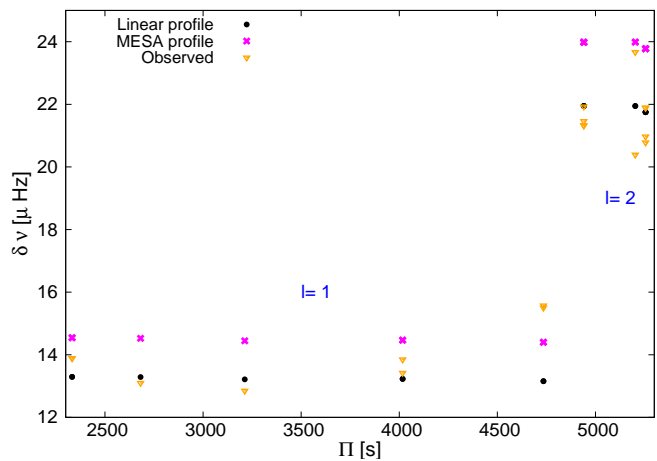} 
\caption{Observed frequency splittings (orange triangles), theoretical frequency splittings obtained from the best-fitting linear differential rotation profile (black dots), and theoretical frequency splittings as resulting from the rotation profile of the {\tt MESA} model (magenta crosses), in terms of the corresponding theoretical periods.}
\label{fig:comparison-splittings}
\end{figure}

In order to compare the main results from the previous sections, we show in Fig.~\ref{fig:comparison-splittings} the observed frequency splittings (orange triangles), the theoretical frequency splittings that result from the use of the best-fitting linear differential profile (black dots) ---in the case where all the observed splittings were fitted, for which we have employed an asteroseismological model representative of GD$\, $278 (generated with {\tt LPCODE} and {\tt LP-PUL} codes).  In addition, magenta crosses indicate the theoretical frequency splittings as predicted from the use of the rotation profile computed with the {\tt MESA} evolutionary code (Table~\ref{tab:compar-splitt}). All these quantities are shown in terms of the corresponding theoretical periods. The figure shows that the theoretical frequency splittings obtained via the simple linear profile and the observed frequency splittings lie close together (being their differences between $\sim 0.02-2.4\ \mu$Hz), while the theoretical frequency splittings obtained from the {\tt MESA} rotation profile lie slightly farther than those observed (differences between $\sim 0.6-3.6\ \mu$Hz). Considering the uncertainties involved in the observed splittings and in the theoretical calculations employed in this work, we believe that these differences are not significant. Thus, we consider that there is a good agreement between the different sets of frequency splittings.

\begin{figure}  
\centering   \includegraphics[clip,width=230pt]{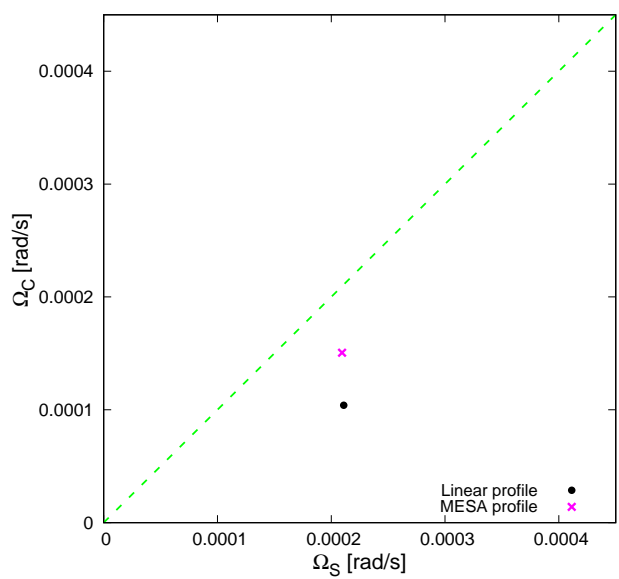} 
\caption{Comparison between the values of $\Omega_S$ and $\Omega_C$ obtained from the best fit solution to the linear differential rotation profiles (black dot) when considering all the splittings with $\ell= 1,2$, and the corresponding values when the rotation profile is that of the {\tt MESA} evolutionary model (magenta cross). The green dashed line indicates the solutions according to rigid rotation.}
\label{fig:comparison-omega}
\end{figure}

Additionally, in Fig.~\ref{fig:comparison-omega} we show the derived values of $\Omega_S$ and $\Omega_C$. In the figure, a black dot indicates the solution corresponding to the best fit in the case of the linear differential rotation profiles, and a magenta cross shows the values of the rotation at the stellar center and surface according to the rotation profile obtained via detailed evolutionary computations with MESA. Again, we indicate the rigid-rotation solutions with a green dashed line. It is clear the agreement between the two approaches, particularly in the value of $\Omega_S$.

Overall, the results from the different approaches
are in line with each other and with the observed properties of GD$\, $278. The close agreement obtained when using a simple artificial rotation profile and the rotation profile that comes from fully evolutionary calculations, strengthens our results, even more given that two independent evolution codes and very different approaches were employed. Our results suggest that the rotation profile is slightly differential but, given the uncertainties involved in these procedures (for instance, uncertainties in the frequency splittings, mode identification, and in the evolutionary approach, mainly in the rotational mixing processes, among others), it is certainly compatible with rigid rotation, as is generally the case for WDs and pre-WDs \citep{2019A&ARv..27....7C}.

\begin{table}
\centering
\caption{Observed and theoretical frequency splittings as predicted by the rotation profile coming from evolutionary calculations.}
\begin{tabular}{cccc}
\hline
 $k$  & $\delta \nu^T$ [$\mu$Hz] & ${\delta \nu^O}$  [$\mu$Hz] & $|\Delta \delta\nu|/\delta\nu$\\ 
\hline
	25	& 14.54		&	13.89	& 0.045		\\
    25   & 14.54    &   13.89   & 0.045                      \\
	29	& 14.52		&	13.10	& 0.098		\\
	35	& 14.44		&	12.85 	& 0.110	 \\
	44	& 14.47		&	13.42 	& 0.072\\
	44	& 14.47	    &	13.85	& 0.043				\\
	52	& 14.40 	&	15.50 	& 0.077	\\
    52	& 14.40 	&	15.57	&0.082\vspace{1.4mm}\\
    
	94	& 23.98 	&	21.32 	& 0.111	\\
    94  & 23.98     &   21.46   & 0.105                      \\
	94	& 23.98		&	21.93	& 0.086		\\
	94	& 23.98		&	21.33 	& 0.111	 \\
	99	& 23.99 	&	20.39	&0.150	\\
    99  & 23.99     &   23.67   & 0.013   \\
	100	& 23.78		&	20.97	& 0.118		\\
	100	& 23.78		&	20.78 	& 0.126	 \\
	100	& 23.78		&	21.89	& 0.079		\\
	100	& 23.78	    &	21.89 	& 0.079	 \\

\hline
\end{tabular}
\label{tab:compar-splitt}
\end{table}

\section{Summary and conclusions}  
\label{conclusions}

In this work, we presented the first exploration of the internal rotation of GD$\, $278, the (for now) only pulsating ELM WD star that exhibits rotational frequency splittings in its periodogram, based on the mode identification presented in 
\cite{2021ApJ...922..220L}. For this purpose, we employed the observed rotational frequency splittings previously identified by \cite{2021ApJ...922..220L} and compared them to theoretical splittings via rotational splitting fits. The theoretical frequency splittings were obtained by considering different \textit{ad hoc} choices of the rotation profiles and employing a stellar model representative of the observed pulsation modes, which is also characterized by stellar properties within the constraints given by the spectroscopic determinations of GD$\, $278 \citep{2020ApJ...889...49B}. This asteroseismological model was obtained by using the {\tt LPCODE} and {\tt LP-PUL} stellar evolution and pulsation codes, respectively. We also followed another approach where we employed a rotation profile resulting from detailed evolutionary calculations obtained with the {\tt MESA} evolution code and used it to infer the expected frequency splittings. 

In order to carry out the rotational splitting fits, we first assumed a rigid body rotation profile, and varied the value of the (constant) angular velocity in a large range of possible values. We obtained that the best fit to all the observed frequency splittings corresponds to an angular velocity of $\sim 1.6 \times 10^{-4}\ $rad s$^{-1}$. Next, we lifted the rigid-body rotation condition, and considered a large set of linear differential rotation profiles, for different values of $\Omega$ at the surface and the center. Again, we took into account all the observed splittings for the fits, and we found that there are multiple possible solutions, \textit{the rigid-body rotation being one of them.} The best-fit solution in this case corresponds to ($\Omega_S,\Omega_C) \sim (2.1 \times 10^{-4}, 1.0 \times 10^{-4}$) rad s$^{-1}$. When we analyzed the rotation profile resulting from the detailed evolutionary calculations with MESA, we found that it is characterized by ($\Omega_S, \Omega_C$) $\sim$ ($2.1 \times 10^{-4}, 1.5 \times 10^{-4}$) rad s$^{-1}$, and comparing these values with the corresponding ones for the best-fitting linear rotation profile, the close agreement is evident, particularly at the surface. These values are only marginally differential and actually compatible with rigid-body rotation. Moreover, a comparison between the observed and the theoretical frequency splittings predicted by both profiles also shows close agreement, being  the theoretical splittings for the \textit{ad hoc} linear case more nearly similar to the observed splittings.

Rotation in ELMVs and pre-ELMVs is relevant since it constitutes a very feasible mechanism candidate to explain the excitation of the pulsations observed in pre-ELM WD stars \citep{2016A&A...585A...1C,2016A&A...595A..35I}, as well as the presence of heavy metals in the surface of ELM WDs \citep{2014ApJ...781..104G,2014MNRAS.444.1674H}. The finding of the first pulsating ELM WD with measurable rotational frequency splittings by \cite{2021ApJ...922..220L} provides, for the first time, the possibility of sounding the internal rotation in this type of stars. In this sense, the present work represents only a first step in the exploration of the internal rotation in ELMVs, since there are still many uncertainties. Indeed, even with state-of-the-art evolution codes and the many advances in the observational field that help the former improve, there is still much work to be done, for instance, in the study of the angular momentum and chemical element transport mechanisms \citep{2017RSOS....470192S}.
The eventual detection of more ELMV stars with measurable rotational splittings can help to place constraints on the physical processes involved in the development of the rotation profiles of these stars, thus making it possible to improve the current stellar models, as in other pulsating stars \citep{2021RvMP...93a5001A,2022ARA&A..60...31K}. For example, few empirical constraints exist on the role of tides in the redistribution of angular momentum at these orbital periods for ELM WDs \citep{2012MNRAS.421..426F,2014MNRAS.444.3488F,2018MNRAS.481..715P}; studies of pulsating hot subdwarfs in close binaries show we have an incomplete understanding of all relevant physics \citep{2019MNRAS.485.2889P}. The future is promising in this regard, given the significant advances that asteroseismology has already benefited from, particularly due to the improvement in the photometry efforts coming from space missions such as Kepler/K2 \citep{2010Sci...327..977B,2014PASP..126..398H} and, more recently, the TESS and CHaracterising ExOPlanets Satellite \citep[Cheops,][]{Moya2018} programs. It is to be expected that future missions like the PLAnetary Transits and Oscillations \citep[Plato,][]{2018EPSC...12..969P}, along with the ongoing TESS and Cheops missions, will bring additional pulsating ELM WDs with detectable rotational splittings and more pulsation measurements.

\begin{acknowledgements}
We wish to thank our anonymous referee for the comments and suggestions that improved the original version of the paper. Support for this work was in part provided by NASA TESS Cycle 2 Grant 80NSSC20K0592 and Cycle 4 grant 80NSSC22K0737. Part of this work was supported by AGENCIA through the Programa de Modernizaci\'on Tecnol\'ogica BID 1728/OC-AR, and by the PIP 112-200801-00940 grant from CONICET. This research has made use of NASA's Astrophysics Data System.
\end{acknowledgements}

\bibliographystyle{aa}
\bibliography{biblio}

\begin{thebibliography}{96}
\expandafter\ifx\csname natexlab\endcsname\relax\def\natexlab#1{#1}\fi

\bibitem[{{Aerts}(2021)}]{2021RvMP...93a5001A}
{Aerts}, C. 2021, Reviews of Modern Physics, 93, 015001

\bibitem[{{Althaus} {et~al.}(2015){Althaus}, {Camisassa}, {Miller Bertolami},
  {C{\'o}rsico}, \& {Garc{\'\i}a-Berro}}]{2015A&A...576A...9A}
{Althaus}, L.~G., {Camisassa}, M.~E., {Miller Bertolami}, M.~M., {C{\'o}rsico},
  A.~H., \& {Garc{\'\i}a-Berro}, E. 2015, \aap, 576, A9

\bibitem[{{Althaus} {et~al.}(2010){Althaus}, {C{\'o}rsico}, {Isern}, \&
  {Garc{\'{\i}}a-Berro}}]{2010A&ARv..18..471A}
{Althaus}, L.~G., {C{\'o}rsico}, A.~H., {Isern}, J., \& {Garc{\'{\i}}a-Berro},
  E. 2010, \aapr, 18, 471

\bibitem[{{Althaus} {et~al.}(2013){Althaus}, {Miller Bertolami}, \&
  {C{\'o}rsico}}]{2013A&A...557A..19A}
{Althaus}, L.~G., {Miller Bertolami}, M.~M., \& {C{\'o}rsico}, A.~H. 2013,
  \aap, 557, A19

\bibitem[{{Althaus} {et~al.}(2009){Althaus}, {Panei}, {Romero}, {Rohrmann},
  {C{\'o}rsico}, {Garc{\'{\i}}a-Berro}, \& {Miller
  Bertolami}}]{2009A&A...502..207A}
{Althaus}, L.~G., {Panei}, J.~A., {Romero}, A.~D., {et~al.} 2009, \aap, 502,
  207

\bibitem[{{Althaus} {et~al.}(2001){Althaus}, {Serenelli}, \&
  {Benvenuto}}]{2001MNRAS.323..471A}
{Althaus}, L.~G., {Serenelli}, A.~M., \& {Benvenuto}, O.~G. 2001, \mnras, 323,
  471

\bibitem[{{Althaus} {et~al.}(2005){Althaus}, {Serenelli}, {Panei},
  {C{\'o}rsico}, {Garc{\'{\i}}a-Berro}, \&
  {Sc{\'o}ccola}}]{2005A&A...435..631A}
{Althaus}, L.~G., {Serenelli}, A.~M., {Panei}, J.~A., {et~al.} 2005, \aap, 435,
  631

\bibitem[{{Bailer-Jones} {et~al.}(2021){Bailer-Jones}, {Rybizki}, {Fouesneau},
  {Demleitner}, \& {Andrae}}]{2021AJ....161..147B}
{Bailer-Jones}, C.~A.~L., {Rybizki}, J., {Fouesneau}, M., {Demleitner}, M., \&
  {Andrae}, R. 2021, \aj, 161, 147

\bibitem[{{Bell} {et~al.}(2019){Bell}, {C{\'o}rsico}, {Bischoff-Kim},
  {Althaus}, {Bradley}, {Calcaferro}, {Montgomery}, {Uzundag}, {Baran},
  {Bogn{\'a}r}, {Charpinet}, {Ghasemi}, \& {Hermes}}]{2019A&A...632A..42B}
{Bell}, K.~J., {C{\'o}rsico}, A.~H., {Bischoff-Kim}, A., {et~al.} 2019, \aap,
  632, A42

\bibitem[{{Bell} {et~al.}(2018){Bell}, {Pelisoli}, {Kepler}, {Brown}, {Winget},
  {Winget}, {Vanderbosch}, {Castanheira}, {Hermes}, {Montgomery}, \&
  {Koester}}]{2018A&A...617A...6B}
{Bell}, K.~J., {Pelisoli}, I., {Kepler}, S.~O., {et~al.} 2018, \aap, 617, A6

\bibitem[{{Bogn{\'a}r} {et~al.}(2014){Bogn{\'a}r}, {Papar{\'o}}, {C{\'o}rsico},
  {Kepler}, \& {Gy{\H{o}}rffy}}]{2014A&A...570A.116B}
{Bogn{\'a}r}, Z., {Papar{\'o}}, M., {C{\'o}rsico}, A.~H., {Kepler}, S.~O., \&
  {Gy{\H{o}}rffy}, {\'A}. 2014, \aap, 570, A116

\bibitem[{{Borucki} {et~al.}(2010){Borucki}, {Koch}, {Basri}, {Batalha},
  {Brown}, {Caldwell}, {Caldwell}, {Christensen-Dalsgaard}, {Cochran},
  {DeVore}, {Dunham}, {Dupree}, {Gautier}, {Geary}, {Gilliland}, {Gould},
  {Howell}, {Jenkins}, {Kondo}, {Latham}, {Marcy}, {Meibom}, {Kjeldsen},
  {Lissauer}, {Monet}, {Morrison}, {Sasselov}, {Tarter}, {Boss}, {Brownlee},
  {Owen}, {Buzasi}, {Charbonneau}, {Doyle}, {Fortney}, {Ford}, {Holman},
  {Seager}, {Steffen}, {Welsh}, {Rowe}, {Anderson}, {Buchhave}, {Ciardi},
  {Walkowicz}, {Sherry}, {Horch}, {Isaacson}, {Everett}, {Fischer}, {Torres},
  {Johnson}, {Endl}, {MacQueen}, {Bryson}, {Dotson}, {Haas}, {Kolodziejczak},
  {Van Cleve}, {Chandrasekaran}, {Twicken}, {Quintana}, {Clarke}, {Allen},
  {Li}, {Wu}, {Tenenbaum}, {Verner}, {Bruhweiler}, {Barnes}, \&
  {Prsa}}]{2010Sci...327..977B}
{Borucki}, W.~J., {Koch}, D., {Basri}, G., {et~al.} 2010, Science, 327, 977

\bibitem[{{Bradley}(2001)}]{2001ApJ...552..326B}
{Bradley}, P.~A. 2001, \apj, 552, 326

\bibitem[{{Brickhill}(1975)}]{1975MNRAS.170..405B}
{Brickhill}, A.~J. 1975, \mnras, 170, 405

\bibitem[{{Brown} {et~al.}(2016){Brown}, {Gianninas}, {Kilic}, {Kenyon}, \&
  {Allende Prieto}}]{2016ApJ...818..155B}
{Brown}, W.~R., {Gianninas}, A., {Kilic}, M., {Kenyon}, S.~J., \& {Allende
  Prieto}, C. 2016, \apj, 818, 155

\bibitem[{{Brown} {et~al.}(2010){Brown}, {Kilic}, {Allende Prieto}, \&
  {Kenyon}}]{2010ApJ...723.1072B}
{Brown}, W.~R., {Kilic}, M., {Allende Prieto}, C., \& {Kenyon}, S.~J. 2010,
  \apj, 723, 1072

\bibitem[{{Brown} {et~al.}(2020){Brown}, {Kilic}, {Kosakowski}, {Andrews},
  {Heinke}, {Ag{\"u}eros}, {Camilo}, {Gianninas}, {Hermes}, \&
  {Kenyon}}]{2020ApJ...889...49B}
{Brown}, W.~R., {Kilic}, M., {Kosakowski}, A., {et~al.} 2020, \apj, 889, 49

\bibitem[{{Brown} {et~al.}(2022){Brown}, {Kilic}, {Kosakowski}, \&
  {Gianninas}}]{2022ApJ...933...94B}
{Brown}, W.~R., {Kilic}, M., {Kosakowski}, A., \& {Gianninas}, A. 2022, \apj,
  933, 94

\bibitem[{{Burgers}(1969)}]{1969fecg.book.....B}
{Burgers}, J.~M. 1969, {Flow Equations for Composite Gases} (New York: Academic
  Press)

\bibitem[{{Calcaferro} {et~al.}(2018{\natexlab{a}}){Calcaferro}, {Althaus}, \&
  {C{\'o}rsico}}]{2018A&A...614A..49C}
{Calcaferro}, L.~M., {Althaus}, L.~G., \& {C{\'o}rsico}, A.~H.
  2018{\natexlab{a}}, \aap, 614, A49

\bibitem[{{Calcaferro} {et~al.}(2016){Calcaferro}, {C{\'o}rsico}, \&
  {Althaus}}]{2016A&A...589A..40C}
{Calcaferro}, L.~M., {C{\'o}rsico}, A.~H., \& {Althaus}, L.~G. 2016, \aap, 589,
  A40

\bibitem[{{Calcaferro} {et~al.}(2017){Calcaferro}, {C{\'o}rsico}, \&
  {Althaus}}]{2017A&A...607A..33C}
{Calcaferro}, L.~M., {C{\'o}rsico}, A.~H., \& {Althaus}, L.~G. 2017, \aap, 607,
  A33

\bibitem[{{Calcaferro} {et~al.}(2018{\natexlab{b}}){Calcaferro}, {C{\'o}rsico},
  {Althaus}, {Romero}, \& {Kepler}}]{2018A&A...620A.196C}
{Calcaferro}, L.~M., {C{\'o}rsico}, A.~H., {Althaus}, L.~G., {Romero}, A.~D.,
  \& {Kepler}, S.~O. 2018{\natexlab{b}}, \aap, 620, A196

\bibitem[{{Cantiello} {et~al.}(2014){Cantiello}, {Mankovich}, {Bildsten},
  {Christensen-Dalsgaard}, \& {Paxton}}]{2014ApJ...788...93C}
{Cantiello}, M., {Mankovich}, C., {Bildsten}, L., {Christensen-Dalsgaard}, J.,
  \& {Paxton}, B. 2014, \apj, 788, 93

\bibitem[{{Charpinet} {et~al.}(2009){Charpinet}, {Fontaine}, \&
  {Brassard}}]{2009Natur.461..501C}
{Charpinet}, S., {Fontaine}, G., \& {Brassard}, P. 2009, \nat, 461, 501

\bibitem[{{C{\'o}rsico} \& {Althaus}(2006)}]{2006A&A...454..863C}
{C{\'o}rsico}, A.~H. \& {Althaus}, L.~G. 2006, \aap, 454, 863

\bibitem[{{C{\'o}rsico} \& {Althaus}(2016)}]{2016A&A...585A...1C}
{C{\'o}rsico}, A.~H. \& {Althaus}, L.~G. 2016, \aap, 585, A1

\bibitem[{{C{\'o}rsico} {et~al.}(2011){C{\'o}rsico}, {Althaus}, {Kawaler},
  {Miller Bertolami}, {Garc{\'\i}a-Berro}, \& {Kepler}}]{2011MNRAS.418.2519C}
{C{\'o}rsico}, A.~H., {Althaus}, L.~G., {Kawaler}, S.~D., {et~al.} 2011,
  \mnras, 418, 2519

\bibitem[{{C{\'o}rsico} {et~al.}(2008){C{\'o}rsico}, {Althaus}, {Kepler},
  {Costa}, \& {Miller Bertolami}}]{2008A&A...478..869C}
{C{\'o}rsico}, A.~H., {Althaus}, L.~G., {Kepler}, S.~O., {Costa}, J.~E.~S., \&
  {Miller Bertolami}, M.~M. 2008, \aap, 478, 869

\bibitem[{{C{\'o}rsico} {et~al.}(2012){C{\'o}rsico}, {Althaus}, {Miller
  Bertolami}, \& {Bischoff-Kim}}]{2012A&A...541A..42C}
{C{\'o}rsico}, A.~H., {Althaus}, L.~G., {Miller Bertolami}, M.~M., \&
  {Bischoff-Kim}, A. 2012, \aap, 541, A42

\bibitem[{{C{\'o}rsico} {et~al.}(2009{\natexlab{a}}){C{\'o}rsico}, {Althaus},
  {Miller Bertolami}, \& {Garc{\'{\i}}a-Berro}}]{2009A&A...499..257C}
{C{\'o}rsico}, A.~H., {Althaus}, L.~G., {Miller Bertolami}, M.~M., \&
  {Garc{\'{\i}}a-Berro}, E. 2009{\natexlab{a}}, \aap, 499, 257

\bibitem[{{C{\'o}rsico} {et~al.}(2009{\natexlab{b}}){C{\'o}rsico}, {Althaus},
  {Miller Bertolami}, {Gonz{\'a}lez P{\'e}rez}, \&
  {Kepler}}]{2009ApJ...701.1008C}
{C{\'o}rsico}, A.~H., {Althaus}, L.~G., {Miller Bertolami}, M.~M.,
  {Gonz{\'a}lez P{\'e}rez}, J.~M., \& {Kepler}, S.~O. 2009{\natexlab{b}}, \apj,
  701, 1008

\bibitem[{{C{\'o}rsico} {et~al.}(2019){C{\'o}rsico}, {Althaus}, {Miller
  Bertolami}, \& {Kepler}}]{2019A&ARv..27....7C}
{C{\'o}rsico}, A.~H., {Althaus}, L.~G., {Miller Bertolami}, M.~M., \& {Kepler},
  S.~O. 2019, \aapr, 27, 7

\bibitem[{{C{\'o}rsico} {et~al.}(2007{\natexlab{a}}){C{\'o}rsico}, {Althaus},
  {Miller Bertolami}, \& {Werner}}]{2007A&A...461.1095C}
{C{\'o}rsico}, A.~H., {Althaus}, L.~G., {Miller Bertolami}, M.~M., \& {Werner},
  K. 2007{\natexlab{a}}, \aap, 461, 1095

\bibitem[{{C{\'o}rsico} {et~al.}(2016){C{\'o}rsico}, {Althaus}, {Serenelli},
  {Kepler}, {Jeffery}, \& {Corti}}]{2016A&A...588A..74C}
{C{\'o}rsico}, A.~H., {Althaus}, L.~G., {Serenelli}, A.~M., {et~al.} 2016,
  \aap, 588, A74

\bibitem[{{C{\'o}rsico} {et~al.}(2007{\natexlab{b}}){C{\'o}rsico}, {Miller
  Bertolami}, {Althaus}, {Vauclair}, \& {Werner}}]{2007A&A...475..619C}
{C{\'o}rsico}, A.~H., {Miller Bertolami}, M.~M., {Althaus}, L.~G., {Vauclair},
  G., \& {Werner}, K. 2007{\natexlab{b}}, \aap, 475, 619

\bibitem[{{C{\'o}rsico} {et~al.}(2021){C{\'o}rsico}, {Uzundag}, {Kepler},
  {Althaus}, {Silvotti}, {Baran}, {Vu{\v{c}}kovi{\'c}}, {Werner}, {Bell}, \&
  {Higgins}}]{2021A&A...645A.117C}
{C{\'o}rsico}, A.~H., {Uzundag}, M., {Kepler}, S.~O., {et~al.} 2021, \aap, 645,
  A117

\bibitem[{{C{\'o}rsico} {et~al.}(2022){C{\'o}rsico}, {Uzundag}, {Kepler},
  {Silvotti}, {Althaus}, {Koester}, {Baran}, {Bell}, {Bischoff-Kim}, {Hermes},
  {Kawaler}, {Provencal}, {Winget}, {Montgomery}, {Bradley}, {Kleinman}, \&
  {Nitta}}]{2022A&A...659A..30C}
{C{\'o}rsico}, A.~H., {Uzundag}, M., {Kepler}, S.~O., {et~al.} 2022, \aap, 659,
  A30

\bibitem[{{Cowling} \& {Newing}(1949)}]{1949ApJ...109..149C}
{Cowling}, T.~G. \& {Newing}, R.~A. 1949, \apj, 109, 149

\bibitem[{{Deheuvels} {et~al.}(2012){Deheuvels}, {Garc{\'\i}a}, {Chaplin},
  {Basu}, {Antia}, {Appourchaux}, {Benomar}, {Davies}, {Elsworth}, {Gizon},
  {Goupil}, {Reese}, {Regulo}, {Schou}, {Stahn}, {Casagrande},
  {Christensen-Dalsgaard}, {Fischer}, {Hekker}, {Kjeldsen}, {Mathur}, {Mosser},
  {Pinsonneault}, {Valenti}, {Christiansen}, {Kinemuchi}, \&
  {Mullally}}]{2012ApJ...756...19D}
{Deheuvels}, S., {Garc{\'\i}a}, R.~A., {Chaplin}, W.~J., {et~al.} 2012, \apj,
  756, 19

\bibitem[{{Fontaine} \& {Brassard}(2008)}]{2008PASP..120.1043F}
{Fontaine}, G. \& {Brassard}, P. 2008, \pasp, 120, 1043

\bibitem[{{Fuller} \& {Lai}(2012)}]{2012MNRAS.421..426F}
{Fuller}, J. \& {Lai}, D. 2012, \mnras, 421, 426

\bibitem[{{Fuller} \& {Lai}(2014)}]{2014MNRAS.444.3488F}
{Fuller}, J. \& {Lai}, D. 2014, \mnras, 444, 3488

\bibitem[{{Gaia Collaboration}(2020)}]{2020yCat.1350....0G}
{Gaia Collaboration}. 2020, VizieR Online Data Catalog, I/350

\bibitem[{{Gentile Fusillo} {et~al.}(2019){Gentile Fusillo}, {Tremblay},
  {G{\"a}nsicke}, {Manser}, {Cunningham}, {Cukanovaite}, {Hollands}, {Marsh},
  {Raddi}, {Jordan}, {Toonen}, {Geier}, {Barstow}, \&
  {Cummings}}]{2019MNRAS.482.4570G}
{Gentile Fusillo}, N.~P., {Tremblay}, P.-E., {G{\"a}nsicke}, B.~T., {et~al.}
  2019, \mnras, 482, 4570

\bibitem[{{Giammichele} {et~al.}(2017){Giammichele}, {Charpinet}, {Brassard},
  \& {Fontaine}}]{2017A&A...598A.109G}
{Giammichele}, N., {Charpinet}, S., {Brassard}, P., \& {Fontaine}, G. 2017,
  \aap, 598, A109

\bibitem[{{Giammichele} {et~al.}(2016){Giammichele}, {Fontaine}, {Brassard}, \&
  {Charpinet}}]{2016ApJS..223...10G}
{Giammichele}, N., {Fontaine}, G., {Brassard}, P., \& {Charpinet}, S. 2016,
  \apjs, 223, 10

\bibitem[{{Gianninas} {et~al.}(2016){Gianninas}, {Curd}, {Fontaine}, {Brown},
  \& {Kilic}}]{2016ApJ...822L..27G}
{Gianninas}, A., {Curd}, B., {Fontaine}, G., {Brown}, W.~R., \& {Kilic}, M.
  2016, \apjl, 822, L27

\bibitem[{{Gianninas} {et~al.}(2014){Gianninas}, {Hermes}, {Brown}, {Dufour},
  {Barber}, {Kilic}, {Kenyon}, \& {Harrold}}]{2014ApJ...781..104G}
{Gianninas}, A., {Hermes}, J.~J., {Brown}, W.~R., {et~al.} 2014, \apj, 781, 104

\bibitem[{{Gianninas} {et~al.}(2015){Gianninas}, {Kilic}, {Brown}, {Canton}, \&
  {Kenyon}}]{2015ApJ...812..167G}
{Gianninas}, A., {Kilic}, M., {Brown}, W.~R., {Canton}, P., \& {Kenyon}, S.~J.
  2015, \apj, 812, 167

\bibitem[{{Heger} {et~al.}(2000){Heger}, {Langer}, \&
  {Woosley}}]{2000ApJ...528..368H}
{Heger}, A., {Langer}, N., \& {Woosley}, S.~E. 2000, \apj, 528, 368

\bibitem[{{Heger} {et~al.}(2005){Heger}, {Woosley}, \&
  {Spruit}}]{2005ApJ...626..350H}
{Heger}, A., {Woosley}, S.~E., \& {Spruit}, H.~C. 2005, \apj, 626, 350

\bibitem[{{Hermes} {et~al.}(2014){Hermes}, {G{\"a}nsicke}, {Koester}, {Bours},
  {Townsley}, {Farihi}, {Marsh}, {Littlefair}, {Dhillon}, {Gianninas},
  {Breedt}, \& {Raddi}}]{2014MNRAS.444.1674H}
{Hermes}, J.~J., {G{\"a}nsicke}, B.~T., {Koester}, D., {et~al.} 2014, \mnras,
  444, 1674

\bibitem[{{Hermes} {et~al.}(2017){Hermes}, {Kawaler}, {Bischoff-Kim},
  {Provencal}, {Dunlap}, \& {Clemens}}]{2017ApJ...835..277H}
{Hermes}, J.~J., {Kawaler}, S.~D., {Bischoff-Kim}, A., {et~al.} 2017, \apj,
  835, 277

\bibitem[{{Hermes} {et~al.}(2013{\natexlab{a}}){Hermes}, {Montgomery},
  {Gianninas}, {Winget}, {Brown}, {Harrold}, {Bell}, {Kenyon}, {Kilic}, \&
  {Castanheira}}]{2013MNRAS.436.3573H}
{Hermes}, J.~J., {Montgomery}, M.~H., {Gianninas}, A., {et~al.}
  2013{\natexlab{a}}, \mnras, 436, 3573

\bibitem[{{Hermes} {et~al.}(2013{\natexlab{b}}){Hermes}, {Montgomery},
  {Winget}, {Brown}, {Gianninas}, {Kilic}, {Kenyon}, {Bell}, \&
  {Harrold}}]{2013ApJ...765..102H}
{Hermes}, J.~J., {Montgomery}, M.~H., {Winget}, D.~E., {et~al.}
  2013{\natexlab{b}}, \apj, 765, 102

\bibitem[{{Hermes} {et~al.}(2012){Hermes}, {Montgomery}, {Winget}, {Brown},
  {Kilic}, \& {Kenyon}}]{2012ApJ...750L..28H}
{Hermes}, J.~J., {Montgomery}, M.~H., {Winget}, D.~E., {et~al.} 2012, \apjl,
  750, L28

\bibitem[{{Howell} {et~al.}(2014){Howell}, {Sobeck}, {Haas}, {Still},
  {Barclay}, {Mullally}, {Troeltzsch}, {Aigrain}, {Bryson}, {Caldwell},
  {Chaplin}, {Cochran}, {Huber}, {Marcy}, {Miglio}, {Najita}, {Smith},
  {Twicken}, \& {Fortney}}]{2014PASP..126..398H}
{Howell}, S.~B., {Sobeck}, C., {Haas}, M., {et~al.} 2014, \pasp, 126, 398

\bibitem[{{Istrate} {et~al.}(2016{\natexlab{a}}){Istrate}, {Fontaine},
  {Gianninas}, {Grassitelli}, {Marchant}, {Tauris}, \&
  {Langer}}]{2016A&A...595L..12I}
{Istrate}, A.~G., {Fontaine}, G., {Gianninas}, A., {et~al.} 2016{\natexlab{a}},
  \aap, 595, L12

\bibitem[{{Istrate} {et~al.}(2016{\natexlab{b}}){Istrate}, {Marchant},
  {Tauris}, {Langer}, {Stancliffe}, \& {Grassitelli}}]{2016A&A...595A..35I}
{Istrate}, A.~G., {Marchant}, P., {Tauris}, T.~M., {et~al.} 2016{\natexlab{b}},
  \aap, 595, A35

\bibitem[{{Itoh} {et~al.}(1987){Itoh}, {Kohyama}, \&
  {Takeuchi}}]{1987ApJ...317..733I}
{Itoh}, N., {Kohyama}, Y., \& {Takeuchi}, H. 1987, \apj, 317, 733

\bibitem[{{Jeffery} \& {Saio}(2013)}]{2013MNRAS.435..885J}
{Jeffery}, C.~S. \& {Saio}, H. 2013, \mnras, 435, 885

\bibitem[{{Kawaler}(2015)}]{2015ASPC..493...65K}
{Kawaler}, S.~D. 2015, in Astronomical Society of the Pacific Conference
  Series, Vol. 493, 19th European Workshop on White Dwarfs, ed. P.~{Dufour},
  P.~{Bergeron}, \& G.~{Fontaine}, 65

\bibitem[{{Kawaler} {et~al.}(1999){Kawaler}, {Sekii}, \&
  {Gough}}]{1999ApJ...516..349K}
{Kawaler}, S.~D., {Sekii}, T., \& {Gough}, D. 1999, \apj, 516, 349

\bibitem[{{Kilic} {et~al.}(2011){Kilic}, {Brown}, {Allende Prieto},
  {Ag{\"u}eros}, {Heinke}, \& {Kenyon}}]{2011ApJ...727....3K}
{Kilic}, M., {Brown}, W.~R., {Allende Prieto}, C., {et~al.} 2011, \apj, 727, 3

\bibitem[{{Kilic} {et~al.}(2018){Kilic}, {Hermes}, {C{\'o}rsico}, {Kosakowski},
  {Brown}, {Antoniadis}, {Calcaferro}, {Gianninas}, {Althaus}, \&
  {Green}}]{2018MNRAS.479.1267K}
{Kilic}, M., {Hermes}, J.~J., {C{\'o}rsico}, A.~H., {et~al.} 2018, \mnras, 479,
  1267

\bibitem[{{Koester} {et~al.}(2009){Koester}, {Voss}, {Napiwotzki},
  {Christlieb}, {Homeier}, {Lisker}, {Reimers}, \&
  {Heber}}]{2009A&A...505..441K}
{Koester}, D., {Voss}, B., {Napiwotzki}, R., {et~al.} 2009, \aap, 505, 441

\bibitem[{{Kosakowski} {et~al.}(2020){Kosakowski}, {Kilic}, {Brown}, \&
  {Gianninas}}]{2020ApJ...894...53K}
{Kosakowski}, A., {Kilic}, M., {Brown}, W.~R., \& {Gianninas}, A. 2020, \apj,
  894, 53

\bibitem[{{Kurtz}(2022)}]{2022ARA&A..60...31K}
{Kurtz}, D.~W. 2022, \araa, 60, 31

\bibitem[{{Kurtz} {et~al.}(2014){Kurtz}, {Saio}, {Takata}, {Shibahashi},
  {Murphy}, \& {Sekii}}]{2014MNRAS.444..102K}
{Kurtz}, D.~W., {Saio}, H., {Takata}, M., {et~al.} 2014, \mnras, 444, 102

\bibitem[{{Ledoux}(1951)}]{1951ApJ...114..373L}
{Ledoux}, P. 1951, \apj, 114, 373

\bibitem[{{Li} {et~al.}(2019){Li}, {Chen}, {Chen}, \&
  {Han}}]{2019ApJ...871..148L}
{Li}, Z., {Chen}, X., {Chen}, H.-L., \& {Han}, Z. 2019, \apj, 871, 148

\bibitem[{{Lopez} {et~al.}(2021){Lopez}, {Hermes}, {Calcaferro}, {Bell},
  {Samuels}, {Vanderbosch}, {C{\'o}rsico}, \& {Istrate}}]{2021ApJ...922..220L}
{Lopez}, I.~D., {Hermes}, J.~J., {Calcaferro}, L.~M., {et~al.} 2021, \apj, 922,
  220

\bibitem[{{Maxted} {et~al.}(2014){Maxted}, {Serenelli}, {Marsh}, {Catal{\'a}n},
  {Mahtani}, \& {Dhillon}}]{2014MNRAS.444..208M}
{Maxted}, P.~F.~L., {Serenelli}, A.~M., {Marsh}, T.~R., {et~al.} 2014, \mnras,
  444, 208

\bibitem[{{Maxted} {et~al.}(2013){Maxted}, {Serenelli}, {Miglio}, {Marsh},
  {Heber}, {Dhillon}, {Littlefair}, {Copperwheat}, {Smalley}, {Breedt}, \&
  {Schaffenroth}}]{2013Natur.498..463M}
{Maxted}, P.~F.~L., {Serenelli}, A.~M., {Miglio}, A., {et~al.} 2013, \nat, 498,
  463

\bibitem[{{Moya} {et~al.}(2018){Moya}, {Barcel\'o Forteza, S.}, {Bonfanti, A.},
  {Salmon, S. J. A. J.}, {Van Grootel, V.}, \& {Barrado, D.}}]{Moya2018}
{Moya}, A., {Barcel\'o Forteza, S.}, {Bonfanti, A.}, {et~al.} 2018, A\&A, 620,
  A203

\bibitem[{{Paxton} {et~al.}(2011){Paxton}, {Bildsten}, {Dotter}, {Herwig},
  {Lesaffre}, \& {Timmes}}]{Paxton2011}
{Paxton}, B., {Bildsten}, L., {Dotter}, A., {et~al.} 2011, \apjs, 192, 3

\bibitem[{{Paxton} {et~al.}(2013){Paxton}, {Cantiello}, {Arras}, {Bildsten},
  {Brown}, {Dotter}, {Mankovich}, {Montgomery}, {Stello}, {Timmes}, \&
  {Townsend}}]{Paxton2013}
{Paxton}, B., {Cantiello}, M., {Arras}, P., {et~al.} 2013, \apjs, 208, 4

\bibitem[{{Paxton} {et~al.}(2015){Paxton}, {Marchant}, {Schwab}, {Bauer},
  {Bildsten}, {Cantiello}, {Dessart}, {Farmer}, {Hu}, {Langer}, {Townsend},
  {Townsley}, \& {Timmes}}]{Paxton2015}
{Paxton}, B., {Marchant}, P., {Schwab}, J., {et~al.} 2015, \apjs, 220, 15

\bibitem[{{Paxton} {et~al.}(2018){Paxton}, {Schwab}, {Bauer}, {Bildsten},
  {Blinnikov}, {Duffell}, {Farmer}, {Goldberg}, {Marchant}, {Sorokina},
  {Thoul}, {Townsend}, \& {Timmes}}]{Paxton2018}
{Paxton}, B., {Schwab}, J., {Bauer}, E.~B., {et~al.} 2018, \apjs, 234, 34

\bibitem[{{Paxton} {et~al.}(2019){Paxton}, {Smolec}, {Schwab}, {Gautschy},
  {Bildsten}, {Cantiello}, {Dotter}, {Farmer}, {Goldberg}, {Jermyn}, {Kanbur},
  {Marchant}, {Thoul}, {Townsend}, {Wolf}, {Zhang}, \& {Timmes}}]{Paxton2019}
{Paxton}, B., {Smolec}, R., {Schwab}, J., {et~al.} 2019, \apjs, 243, 10

\bibitem[{{Pelisoli} {et~al.}(2018){Pelisoli}, {Kepler}, {Koester},
  {Castanheira}, {Romero}, \& {Fraga}}]{2018MNRAS.478..867P}
{Pelisoli}, I., {Kepler}, S.~O., {Koester}, D., {et~al.} 2018, \mnras, 478, 867

\bibitem[{{Piotto}(2018)}]{2018EPSC...12..969P}
{Piotto}, G. 2018, in European Planetary Science Congress, EPSC2018--969

\bibitem[{{Preece} {et~al.}(2018){Preece}, {Tout}, \&
  {Jeffery}}]{2018MNRAS.481..715P}
{Preece}, H.~P., {Tout}, C.~A., \& {Jeffery}, C.~S. 2018, \mnras, 481, 715

\bibitem[{{Preece} {et~al.}(2019){Preece}, {Tout}, \&
  {Jeffery}}]{2019MNRAS.485.2889P}
{Preece}, H.~P., {Tout}, C.~A., \& {Jeffery}, C.~S. 2019, \mnras, 485, 2889

\bibitem[{{Ricker} {et~al.}(2015){Ricker}, {Winn}, {Vanderspek}, {Latham},
  {Bakos}, {Bean}, {Berta-Thompson}, {Brown}, {Buchhave}, {Butler}, {Butler},
  {Chaplin}, {Charbonneau}, {Christensen-Dalsgaard}, {Clampin}, {Deming},
  {Doty}, {De Lee}, {Dressing}, {Dunham}, {Endl}, {Fressin}, {Ge}, {Henning},
  {Holman}, {Howard}, {Ida}, {Jenkins}, {Jernigan}, {Johnson}, {Kaltenegger},
  {Kawai}, {Kjeldsen}, {Laughlin}, {Levine}, {Lin}, {Lissauer}, {MacQueen},
  {Marcy}, {McCullough}, {Morton}, {Narita}, {Paegert}, {Palle}, {Pepe},
  {Pepper}, {Quirrenbach}, {Rinehart}, {Sasselov}, {Sato}, {Seager},
  {Sozzetti}, {Stassun}, {Sullivan}, {Szentgyorgyi}, {Torres}, {Udry}, \&
  {Villasenor}}]{2015JATIS...1a4003R}
{Ricker}, G.~R., {Winn}, J.~N., {Vanderspek}, R., {et~al.} 2015, Journal of
  Astronomical Telescopes, Instruments, and Systems, 1, 014003

\bibitem[{{Ritter}(1988)}]{Ritter1988}
{Ritter}, H. 1988, \aap, 202, 93

\bibitem[{{Romero} {et~al.}(2012){Romero}, {C{\'o}rsico}, {Althaus}, {Kepler},
  {Castanheira}, \& {Miller Bertolami}}]{2012MNRAS.420.1462R}
{Romero}, A.~D., {C{\'o}rsico}, A.~H., {Althaus}, L.~G., {et~al.} 2012, \mnras,
  420, 1462

\bibitem[{{Salaris} \& {Cassisi}(2017)}]{2017RSOS....470192S}
{Salaris}, M. \& {Cassisi}, S. 2017, Royal Society Open Science, 4, 170192

\bibitem[{{Spruit}(2002)}]{2002A&A...381..923S}
{Spruit}, H.~C. 2002, \aap, 381, 923

\bibitem[{{Sun} \& {Arras}(2018)}]{2018ApJ...858...14S}
{Sun}, M. \& {Arras}, P. 2018, \apj, 858, 14

\bibitem[{{Unno} {et~al.}(1989){Unno}, {Osaki}, {Ando}, {Saio}, \&
  {Shibahashi}}]{1989nos..book.....U}
{Unno}, W., {Osaki}, Y., {Ando}, H., {Saio}, H., \& {Shibahashi}, H. 1989,
  {Nonradial oscillations of stars}

\bibitem[{{Uzundag} {et~al.}(2022){Uzundag}, {C{\'o}rsico}, {Kepler},
  {Althaus}, {Werner}, {Reindl}, \& {Vu{\v{c}}kovi{\'c}}}]{2022MNRAS.513.2285U}
{Uzundag}, M., {C{\'o}rsico}, A.~H., {Kepler}, S.~O., {et~al.} 2022, \mnras,
  513, 2285

\bibitem[{{Van Grootel} {et~al.}(2013){Van Grootel}, {Fontaine}, {Brassard}, \&
  {Dupret}}]{2013ApJ...762...57V}
{Van Grootel}, V., {Fontaine}, G., {Brassard}, P., \& {Dupret}, M.-A. 2013,
  \apj, 762, 57

\bibitem[{{Wang} {et~al.}(2020){Wang}, {Zhang}, \& {Dai}}]{2020ApJ...888...49W}
{Wang}, K., {Zhang}, X., \& {Dai}, M. 2020, \apj, 888, 49

\bibitem[{{Winget} \& {Kepler}(2008)}]{2008ARA&A..46..157W}
{Winget}, D.~E. \& {Kepler}, S.~O. 2008, \araa, 46, 157

\end{thebibliography}

\end{document}